\input{aipcheck}

\documentclass[
final            
  ]
  {aipproc}

\layoutstyle{8x11double}

\renewcommand\XFMtitleblock{%
  \XFMtitle
  \let\XFMoldpar\par
  \def\par{\XFMoldpar\def\par{\space 
             (on behalf of the H.E.S.S. Collaboration)\XFMoldpar}}%
   \XFMauthors
   \let\par\XFMoldpar
   \XFMaddresses
   \XFMabstract
   \vspace{5pt}%
   \XFMkeywords
   \XFMclassification
 }

\begin{document}

\title{Very-High Energy Gamma-Ray Flux Limits for Nearby Active Galactic Nuclei
}


\author{T. Herr}{
  address={Max-Planck-Institut f\"ur Kernphysik, Saupfercheckweg 1, 69117 Heidelberg, Germany}
}

\author{W. Hofmann}{
  address={Max-Planck-Institut f\"ur Kernphysik, Saupfercheckweg 1, 69117 Heidelberg, Germany}
}

\begin{abstract}
Combining the results of targeted observations, H.E.S.S. has accumulated a large amount of extra-galactic exposure at TeV energies. Due to its large field of view a considerable part of the sky ($0.6$ sr) has been observed with high sensitivity outside the targeted observation positions. Since this exposure region contains little inherent bias, it is well suited for studies of extra-galactic source populations. Given the correlation between ultra-high energy cosmic rays and nearby extra-galactic objects recently claimed by the Auger collaboration, this unbiased sky sample by H.E.S.S. is of interest since it includes (besides the targeted sources) $63$ AGN within $100$~Mpc, for which very-high energy gamma-ray flux limits are derived.
\end{abstract}

\maketitle


\section{Introduction}
Active galactic nuclei (AGN) have long been considered as sites where acceleration of particles to the highest observed energies might be possible \cite{Hillas84}. Recent claims by the Auger collaboration of a correlation between ultra-high energy cosmic ray (UHECR) arrival directions and nearby AGN positions have further strengthened this conjecture \cite{Auger07Sci}. Although AGN can not unambiguously be identified as the sources of the observed UHECRs, they are prime candidates within the Greisen-Zatsepin-Kuzmin cutoff distance. Point-like (for H.E.S.S.) very-high energy (VHE) gamma-ray emission could be produced at the acceleration sites or outside the UHECR sources in electromagnetic cascades \cite{Ferrigno05} or synchrotron processes \cite{Gabici07}.

\section{Analysis}
Due to its large field of view of $5^{\circ}$, H.E.S.S. covers an additional sky area in each observation besides the actual observation target. Assuming that the distribution of sources contained in the additional sky area is independent of the observation target, this additional field is unbiased and free of selection effects and therefore suited for statistical analyses. All H.E.S.S. data taken in targeted observations are combined to derive all-sky maps providing significance and flux upper limits for VHE gamma-ray signals for any sky position covered by H.E.S.S.. These maps are then used to investigate a possible correlation of nearby AGN and point-like VHE gamma-ray emission.\\
This section describes the data set, the production of significance and VHE flux upper limit maps and the selection of AGN. Preliminary results and conclusions are presented in the sections thereafter.

\subsection{Data Set}
\begin{figure}[htbp]
  \includegraphics[width=.9\textwidth]{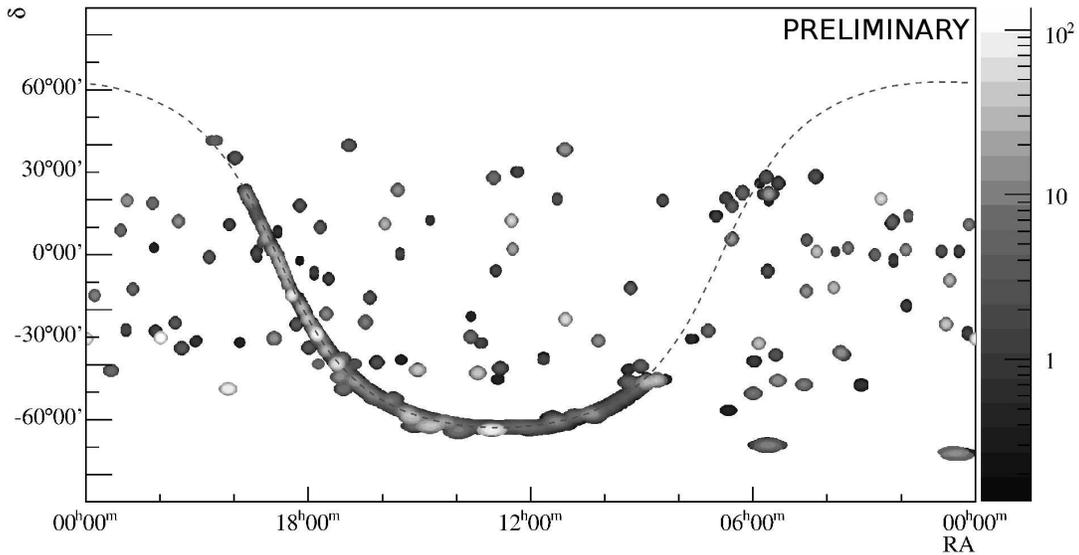}
  \caption{All-sky map of the H.E.S.S. exposure given by the gray scale in hours of equivalent targeted observation time. The dashed line represents the galactic plane which appears curved in the used RA-Dec (J2000) coordinate system (RA, $\mathrm{\delta}$). The observation time splits roughly equally in galactic and extra-galactic pointings. }
\label{acct}
\end{figure}

The data set comprises all H.E.S.S.
observations from March 2004 through December 2007. 
Only observations where at least three of the four 
telescopes participated in data taking are considered. 
After quality selection this corresponds to roughly 
$2800$~hours of observation time, half on
extra-galactic targets. Fig. \ref{acct} gives an impression of the sky areas covered by H.E.S.S. 
and the respective exposure. Here and in the following exposure 
always refers to the equivalent targeted observation time.

\subsection{Significance and Flux Limit Maps}
The data are reduced using the
H.E.S.S. standard ana\-ly\-sis tools and selection
cuts (standard cuts) \cite{Crab06}. Reconstructed 
gamma-like events within $2^{\circ}$ of the respective
observation position are then accumulated in all-sky maps.
The cosmic ray background and the number of excess counts is estimated 
at any covered sky position using the ring-background 
model for point sources \cite{Berge07}. Significances are calculated following 
the prescription of Li \& Ma \cite{LiMa83}. Upper limits on the number of 
excess counts are calculated at $99.9\%$~c.l. using the method of Feldman \& Cousins \cite{FC98}.
These calculations are based on the number of on-and off-counts and the ratio $\alpha$ between the  
weighted areas of on- and off-regions, where the weighting accounts for the different and varying acceptance for background
events in these regions.\\
Upper limits on the integral VHE gamma-ray flux are derived by relating the upper limit on the number of observed excess counts to the number of excess counts that would be expected from an assumed VHE gamma-ray reference source of defined integral flux and spectral shape. The spectrum of the reference source is taken to be a power law with a spectral index of~3 which is assumed to match the unknown AGN spectra. The number of expected excess counts is given by
 the time integrated, energy dependent\footnote{
The effective detection area also depends on the angular distance to the telescopes' pointing 
direction and the zenith angle of the observation. On longer time scales it is also affected by degradation 
of the telescopes' mirrors.} 
effective detection area of H.E.S.S.
convolved with the spectrum of the reference source.
Dividing the upper limit on the number of observed excess counts by 
the number of expected excess counts and multiplying with the previously
 defined integral flux of the reference source eventually gives the upper limit on 
integral flux for the observed sky position.\\
Significance and flux upper limit at a given AGN position can then 
readily be read off the respective all-sky map.

\subsection{Selection of AGN}
\begin{figure}[htbp]
  \includegraphics[width=.9\textwidth]{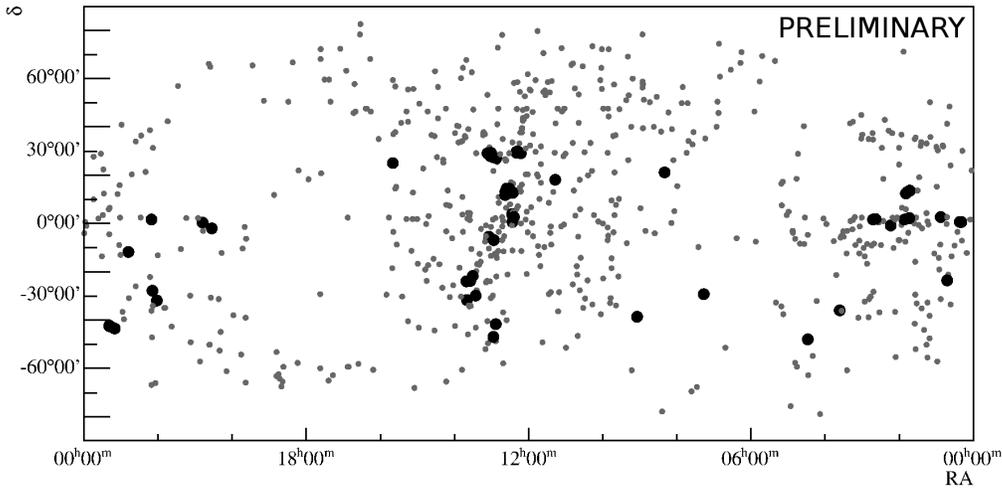}
  \caption{Position of $694$ AGN within $100$~Mpc (gray circles) taken from the 12$^{\mathrm{th}}$~edition of the V\'eron-Cetty and V\'eron Catalog of Quasars and Active galactic nuclei. The map corresponds to the one shown in Fig. \ref{acct}. AGN covered by H.E.S.S. are marked with larger black circles.}
  \label{agn}
\end{figure}

Positions and distances of
AGN closer than $100 \mathrm{~Mpc}$ (redshift $z < 0.024$) are taken
from the 12$^{\mathrm{th}}$~edition of the V\'eron-Cetty and
V\'eron Catalog of Quasars and Active Galactic
Nuclei \cite{VeronCetty06}. Only AGN are considered for which flux upper limits on the order of $10\%$ or less of the
Crab nebula flux\footnote{In this analysis an integral VHE gamma-ray-flux of $2.19 \times 10^{-11} \mathrm{~cm}^{-2} \mathrm{~s}^{-1}$ above $1$~TeV is found for the Crab nebula assuming a power law spectrum with index $2.63$. This agrees well with the value found in \cite{Crab06}.} (Crab unit) can be expected, based on
their exposure.
 To minimize statistical bias,
targeted objects are excluded.
Objects that are too close to regions of known VHE gamma-ray emission
 or bright stars are also excluded to
avoid systematic effects. The final sample
consists of $63$ AGN which is considered to be an
essentially unbiased fraction of the $694$
AGN within $100$~Mpc (see Fig. \ref{agn}).

\section{Results}
\begin{figure}[htbp]
  \includegraphics[width=.45\textwidth]{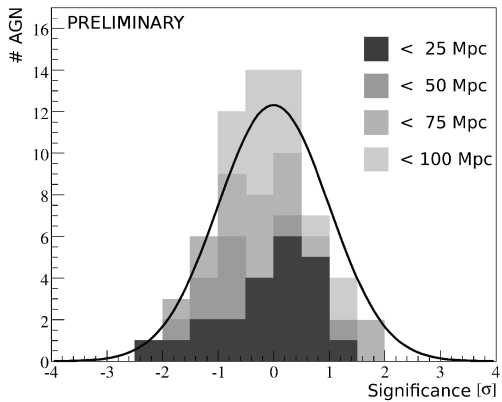}
  \caption{Significance distribution (in units of the standard deviation $\sigma$) of the VHE gamma-ray excess of $63$ AGN within $25$, $50$, $75$ or $100$~Mpc distance. The black line corresponds to a Gaussian distribution of zero mean and unit width.}
  \label{sigall}
\end{figure}

\begin{figure}[htbp]
  \includegraphics[width=.45\textwidth]{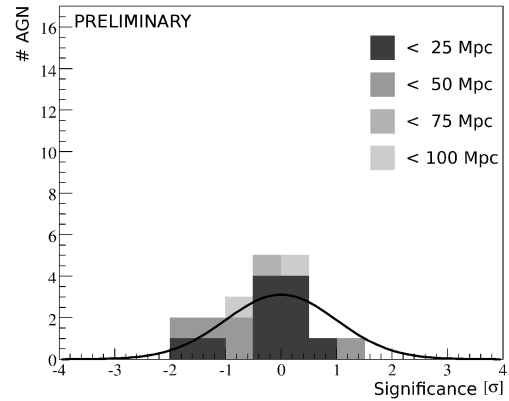}
  \caption{Same as Fig. \ref{sigall}, for the subset of $19$ AGN with the highest exposure.}
  \label{sighigh}
\end{figure}

\begin{figure}[htbp]
  \includegraphics[width=.45\textwidth]{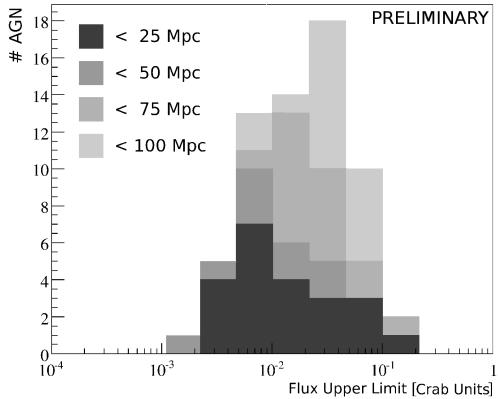}
  \caption{$99.9\%$~c.l. upper limits on the integral flux above $1$~TeV for $63$ AGN within $25$, $50$, $75$ or $100$~Mpc distance. A spectral index of $3.0$ was assumed. }
  \label{fluxul}
\end{figure}

All results presented here are still preliminary. They are based on the values read off the significance and flux limit maps at the respective AGN position.\\
Fig. \ref{sigall} shows the significance distribution of the $63$ selected nearby AGN in the H.E.S.S. field of view. The significance distribution of a subset of $19$ AGN with the highest exposure (minimum exposure for selection increased by a factor $100$) can be seen in Fig. \ref{sighigh}. Neither one of the two distributions shows evidence for a positive correlation between positions of nearby AGN and VHE gamma-ray emission.\\
Integral VHE gamma-ray flux limits above $1$~TeV at $99.9\%$~c.l. for all $63$ AGN in the sample are shown in Fig. \ref{fluxul}. While probably not suitable for detection (too many different AGN subclasses) a stacking approach can be used to constrain the average VHE gamma-ray flux of nearby AGN. The stacked exposure of the $63$ AGN corresponds to more than $550$ hours of H.E.S.S. targeted observations. Based on the stacked values for on-counts, off-counts, $\alpha$ and expected reference excess counts (see previous section) a stacked significance of $S = -1.29 \mathrm{~\sigma}$ and a stacked upper limit on the integral flux above $1$~TeV of $\Phi(>1TeV) < 2.7 \times 10^{-3} \mathrm{~Crab~units}$ at $99.9\%$~c.l. is obtained for the AGN sample. For a spectral index of $3$ this translates into an energy flux limit of roughly $F(>1TeV) < 1.9 \times 10^{-13}\mathrm{~erg~cm}^{-2}\mathrm{~s}^{-1}$. The average (exposure weighted) inverse squared distance of the AGN sample is approximately $d^{-2}=(10\mathrm{~Mpc})^{-2}$. 
Assuming equally bright and isotropically emitting sources with a generic VHE gamma-ray luminosity $L$, an upper limit on this average luminosity above $1$~TeV can be derived:\\

  $ L(>1TeV) = F(>1TeV) \times 4 \pi d^2 < 2.3 \times 10^{39} \mathrm{~erg~s}^{-1}$ \\

\section{Conclusions}
H.E.S.S. extra-galactic fields contain an essentially unbiased sample of $63$ non-targeted AGN 
within $100$~Mpc for which preliminary results are presented. 
There is no evidence for a correlation between nearby AGN and VHE gamma-ray emission as seen by H.E.S.S.. 
Flux limits on the order of $10\%$ of the Crab nebula flux or
better are derived. A stacking approach can be used to constrain the average VHE gamma-ray emission of nearby AGN.
All values were taken from all-sky sensitivity and flux maps, which in general could provide a powerful tool
for future correlation and stacking ana\-ly\-ses of all kinds of extra-galactic source populations.





\bibliographystyle{aipproc}   

\bibliography{G08Bib}

\IfFileExists{\jobname.bbl}{}
 {\typeout{}
  \typeout{******************************************}
  \typeout{** Please run "bibtex \jobname" to obtain}
  \typeout{** the bibliography and then re-run LaTeX}
  \typeout{** twice to fix the references!}
  \typeout{******************************************}
  \typeout{}
 }

\end{document}